**Direct-write milling of diamond by a focused oxygen ion beam**

Aiden A. Martin,[1] Steven Randolph,[2] Aurelien Botman,[2] Milos Toth,[1, a)] and Igor Aharonovich[1, b)]

1) School of Physics and Advanced Materials, University of Technology, Sydney, 15 Broadway, Ultimo, New South Wales 2007, Australia

2) FEI Company, 5350 Northeast Dawson Creek Drive, Hillsboro, Oregon 97214-5793, USA

Recent advances in focused ion beam technology have enabled high-resolution, direct-write nanofabrication using light ions. Studies with light ions to date have, however, focused on milling of materials where sub-surface ion beam damage does not inhibit device performance. Here we report on direct-write milling of single crystal diamond using a focused beam of oxygen ions. Material quality is assessed by Raman and luminescence analysis, and reveals that the damage layer generated by oxygen ions can be removed by non-intrusive post-processing methods such as localised electron beam induced chemical etching.

a) Electronic mail: Milos.Toth@uts.edu.au
b) Electronic mail: Igor.Aharonovich@uts.edu.au

Focused ion beam (FIB) milling is a popular technique for rapid, direct-write nanofabrication via the sputtering of target material through momentum transfer from an energetic primary ion[1]. Most commercial FIB systems are equipped with gallium liquid metal ion sources[2]. However, advances made over the past decade in technologies such as gas field ion[3-5] and inductively coupled plasma[6,7] sources (ICPS) have increased the use of light species (H, He, O and Ne) in the fabrication of nanostructures[8-10]. These species enable control over implantation, sputtering and chemical interactions with the target, but the damage generated by many of these ions is poorly understood at the high fluencies required for FIB milling[6,11] (relative to those used for implantation[12,13]).

In this letter, we report on nanoscale, direct-write milling of optical structures in single crystal diamond using a focused beam of oxygen ions. To characterise the influence of the ions on material properties, we employ photoluminescence (PL) and Raman spectroscopy, and constant power cathodoluminescence (CL) depth profiling of the nitrogen-vacancy (NV) colour centre[14]. We find the neutral nitrogen-vacancy ($NV^0$) CL emission is quenched over depths much greater than those expected from simple binary approximation Monte Carlo simulations of ion interactions with matter[15]. We conclude that oxygen channelling contributes significantly to the sub-surface damage profile of single crystal diamond. The damage layer is significantly thicker than that generated by $Ga^+$ ions. It can, however, be removed by relatively non-intrusive localised methods such as $H_2O$-mediated electron beam induced etching[1,16,17] (EBIE).

Diamond is a promising material for quantum photonic applications due to it's unique chemical, physical and optical properties[18]. During FIB milling, momentum transfer from the ions to the target atoms above a critical dose[11,19] induces amorphisation of diamond within the ion interaction volume. For gallium based FIB milling this amorphisation damage profile is on the order of 46 nm at an ion energy of 30 keV, and the damage layer is heavily stained by implanted gallium ions11. It has been shown that gallium staining can be partly removed using a hydrogen plasma and chemical etch treatment[20,21]. In general, however, it is desirable to utilise in situ techniques such as EBIE, which are typically free from material degradation caused by material incompatibilities with aggressive chemical treatments.

An undercut bridge structure[20] was fabricated in (100) oriented single crystal diamond (CVD grown, Element Six) by a focused oxygen beam ($O^+$ / $O_2^+$ ions, approximately 1:1 relative ion abundance ratio) using an FEI Vion FIB[22] modified to incorporate an O2 source. The structure was fabricated by milling two 25 × 5 µm boxes separated by 2 µm using a 30 kV oxygen ion beam incident normal to the diamond. The gap was then undercut by milling at

45° until visual endpoint, rotating the sample through 180° and then milling further until visual endpoint. The resulting free-standing diamond structure is shown in the inset of Figure 1a.

The bridge and non-processed diamond were analysed by Raman spectroscopy using a Renishaw in Via Raman microscope with a 633 nm excitation laser (Figure 1a). The bridge exhibits a broad Raman scattering profile, similar to that found in amorphous carbon[23] (we note that the diamond line at 1332 cm$^{-1}$ originates from non-processed diamond located underneath the ion-induced damage layer[24]). Figure 1b shows room temperature PL spectra of non-processed diamond and the bridge obtained using a custom confocal microscope with a 532 nm excitation laser. PL from the bridge is characterised by significant quenching of the negative nitrogen-vacancy (NV$^-$) emission, an intense broad-band emission that overlaps with the NV$^-$ zero phonon line (ZPL), and the neutral vacancy defect `GR1' ZPL which is characteristic of ion bombarded diamond[25].

To determine the range of damage in diamond caused by oxygen FIB milling, a series of boxes were fabricated at normal incidence. Boxes were fabricated with ion energies of 2, 8, 16 and 30 keV (and fluencies of 5, 27, 27 and 27 nC/µm$^2$ respectively). Constant power CL depth profiling[26,27] of the NV$^0$ colour centre was used to examine the ion induced damage layers using a FEI Quanta 200 scanning electron microscope (SEM) equipped with a Gatan parabolic CL collection mirror. The mirror was positioned above the sample and directed the emitted light to a Hamamatsu charge-coupled device (CCD). Conveniently, the NV0 emission probed by CL does not overlap with the additional defect-related emissions seen in PL spectra (Figure 1b), as illustrated by the CL spectrum shown in Figure 1c.

Figure 2a shows NV$^0$ CL depth profiles obtained from non-processed diamond and regions milled by 2, 8, 16 and 30 keV oxygen ions. The NV0 emission intensity was measured as a function of electron beam energy using an electron beam power of 40 µW, and a 10 nm bandpass centered on 575 nm (see Figure 1c). To correlate the electron beam energy with the maximum CL generation depth, the electron energy deposition profiles shown in Figure 2b were simulated using standard Monte Carlo models[28] of electron-solid interactions. The curves in Figure 2b are approximately proportional to CL generation profiles within the electron interaction volume[26] in diamond and show how the maximum CL generation depth scales with beam energy.

The CL depth profiles in Figure 2a show that the NV$^0$ emission is quenched within a near surface region whose thickness increases with the energy of the oxygen ions. To quantify the thickness of this damaged region, we used the x-intercept of each dataset in Figure 2a as a measure of the electron beam energy that corresponds to the onset of NV0 emission. At energies lower than the onset, the CL generation volume is contained within the damaged near-surface region where the $NV^0$ emission is quenched. For example, in the case of diamond milled by 30 keV oxygen ions, the $NV^0$ emission onset is observed at an electron beam energy of 8 keV, which corresponds to a maximum CL generation depth of 500 nm (see Figure 2b). Hence, the first 500 nm of the diamond is comprised of damaged, non-luminescent material. This method was applied to all four regions irradiated by oxygen ions to determine the damage range, shown in the inset of Figure 2b, as a function of oxygen ion energy. We note that `non-processed' diamond also displays quenching of the $NV^0$ emission within the first ~ 50 nm of surface material. We ascribe this damage layer to non-luminescent defects produced by unfocused oxygen neutrals that bombard the sample during FIB processing.

The range of damage generated by oxygen ions (inset of Figure 2b) is significantly greater than that produced by gallium ions (e.g. at 30 keV, oxygen ions give rise to a damage range of 500 nm, whereas Ga+ FIB milling of diamond has been observed experimentally to generate a damage layer thickness of 46 nm[11]). To determine the reason behind the large damage range of oxygen ions, we compare the CL depth profiles to the range of O+ ions in amorphous carbon simulated using SRIM[15] and the properties of diamond (density = 3.515 g cm$^{-3}$, displacement energy = 40 eV[29]).

Oxygen ion implantation depth distributions simulated for ion energies of 2, 8, 16 and 30 keV are shown in Figure 2c. Here we plot the range of oxygen ion implantation and not the range of vacancies induced by ions as we cannot exclude the possibility of an oxygen related non-radiative centre quenching NV emission. The two ranges however, are nearly identical in this ion energy range. At 30 keV, O ions penetrate to a depth of ~ 60 nm in amorphous carbon material, which is significantly smaller than the damage range of 500 nm found by CL depth profiling of single crystal diamond. We ascribe this difference between SRIM and experimental results primarily to channelling of O ions in the single crystal diamond structure (which is neglected by the binary approximation model implemented in SRIM). Swelling of the material due to ion implantation[30] is a potential secondary cause of the thick damage layers. At the ion energies used here, channelling of 15N in diamond displays a similar difference between simulated and experimental data[31]. Channelling can be minimized by sample tilting. However, this is not optimal for the fabrication of arbitrary structure geometries, and the effectiveness and applicability of this method decreases with decreasing ion energy[32] and ion mass[33,34] due to corresponding increases in the critical angles for channelling.

Next, we turn to post-fabrication removal of the damage layer generated by oxygen ions. Previously, some methods have been shown to remove amorphous material from diamond damaged by $Ga^+$ FIB milling[20,21]. The ultimate goal of such post-processing treatments is complete, localised removal of the damaged layer and impurities implanted by the ion beam without the need for harsh chemical treatments that can be detrimental to hybrid diamond-based devices. $H_2O$-mediated EBIE is a nanoscale, localised dry chemical etch technique[1,16] that volatilises carbon, and does not compromise the luminescent properties of single crystal diamond17 (currently, no EBIE method exists for the removal of implanted gallium impurities). We applied $H_2O$-mediated EBIE to a region of diamond that had been milled by 30 keV focused oxygen ions. EBIE was performed at room temperature using a Quanta 200 variable pressure[35] SEM, a $H_2O$ pressure of 100 Pa, and a focused, 25 keV, 8.3 nA electron beam scanned over an area of 5 × 10 _m for 45 minutes (primary electron fluence = 2.8 × $10^{20}$ $cm^{-2}$). The milled area was imaged ex situ using the tapping mode of a DI Dimension 3100 atomic force microscope (AFM), and analysed using the software package Gwyddion36 which shows that ~ 70 nm of material was removed by EBIE (Figure 3).

CL depth profiles were obtained from the oxygen FIB milled and EBIE polished regions. The region etched by EBIE displays greatly enhanced CL emission when compared to the as milled region (Figure 3). While only 70 nm of material was removed from the 500 nm damage layer, the CL depth profile from the resulting region yields a damage layer thickness of ~ 250 nm. We tentatively ascribe this discrepancy to partial recovery of material swelling and of damage caused by ion channelling by the high fluence electron beam irradiation treatment used for EBIE. Qualitatively, our results show that EBIE does not induce further amorphisation of the underlying pristine diamond and that it is a viable technique for removing damage layers generated by oxygen ions.

In summary, we have characterised the damage induced in single crystal (100) diamond during direct-write milling with a focused oxygen ion beam. The thickness of a damage layer in which NV centre luminescence is quenched is shown to be significantly greater than the oxygen implantation range predicted by Monte Carlo binary approximation models. We ascribe the difference to channelling and volume expansion of the milled region. The damage layer can be removed by $H_2O$-mediated electron beam induced etching of carbon.

## ACKNOWLEDGEMENTS

This work was supported by FEI Company and the Australian Research Council (Project Number DP140102721). A.A.M. is the recipient of a John Stocker Postgraduate Scholarship from the Science and Industry Endowment Fund. I.A. is the recipient of an Australian Research Council Discovery Early Career Research Award (Project Number DE130100592).

## AUTHOR CONTRIBUTIONS

A.A.M., S.R., A.B., M.T. and I.A. designed the project. A.A.M., S.R., A.B. and I.A. performed the experiments. All authors analysed the data and contributed to the writing of the manuscript.

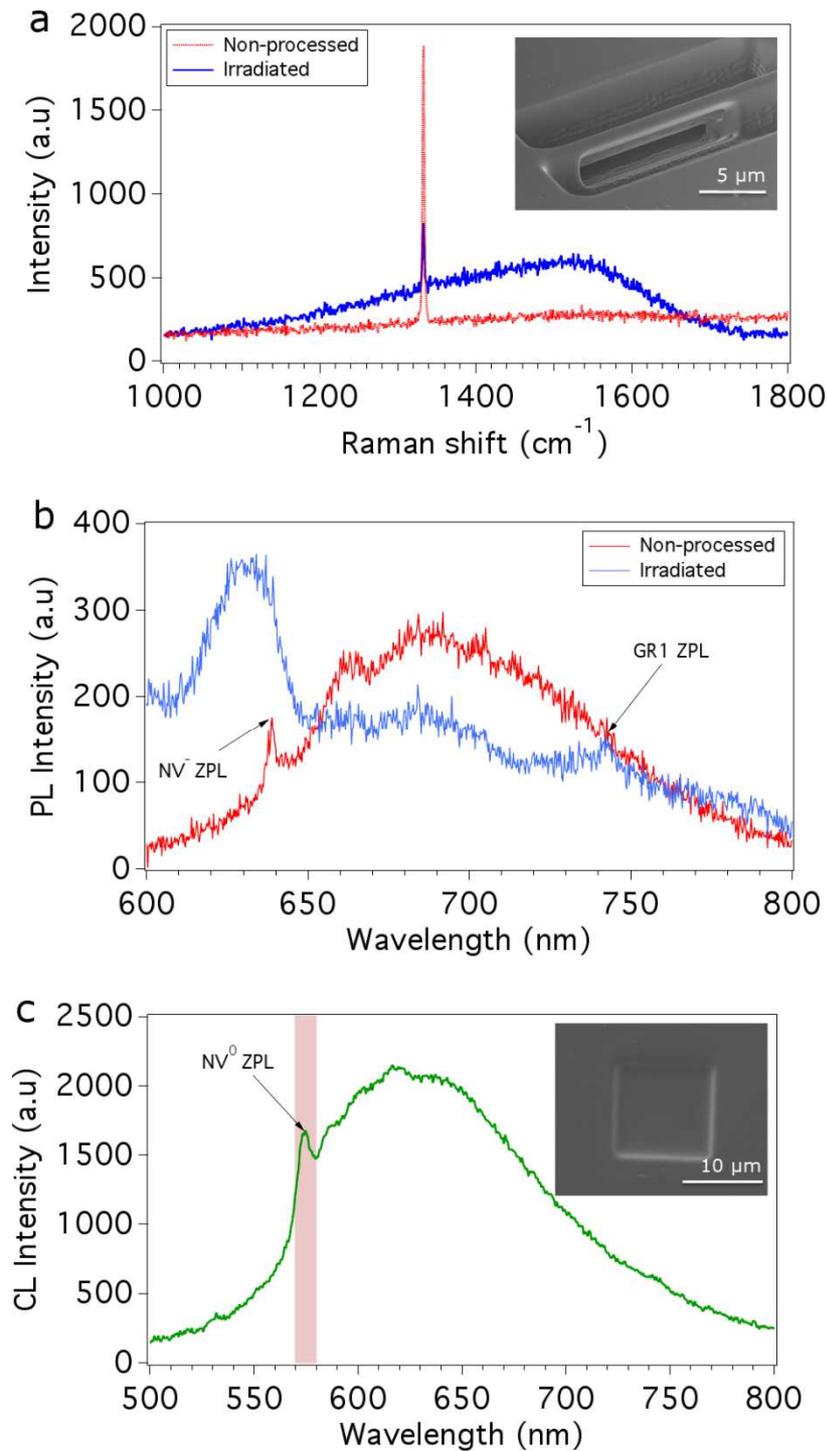

Figure 1. **Optical quality of diamond structures fabricated using a focused oxygen ion beam.** a) Raman spectra of non-processed diamond and a bridge structure (shown in the inset) fabricated using 30 keV oxygen ions. b) PL spectra of non-processed diamond and the bridge. c) CL spectrum obtained using a 15 keV electron beam of a region of diamond (shown in the inset) irradiated by 30 keV oxygen ions.

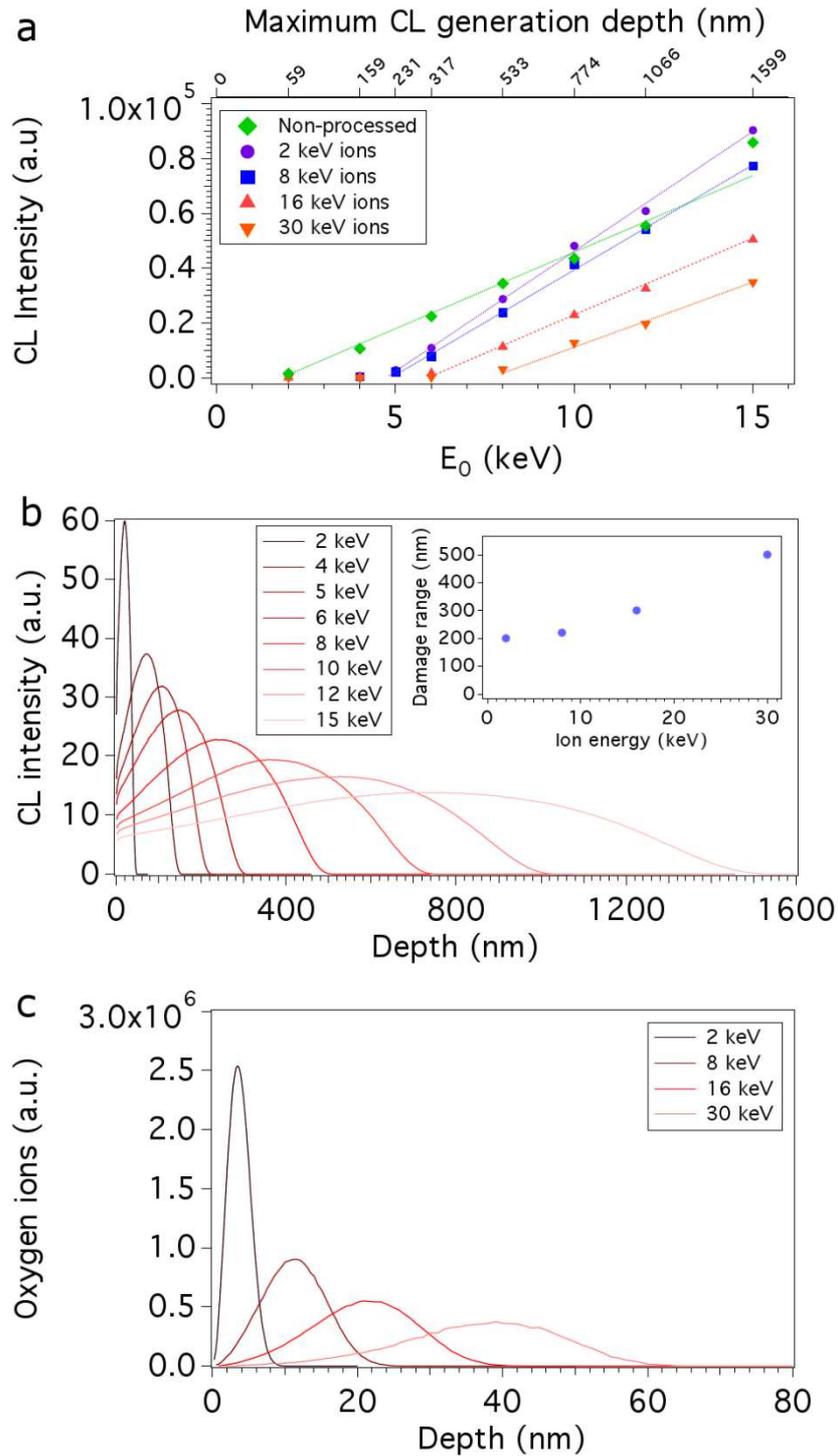

Figure 2. **Profiling of oxygen ion induced damage in diamond.** a) CL depth profiles measured from non-processed single crystal diamond and regions milled by 2, 8, 16 and 30 keV oxygen ions. b) CL generation profiles simulated for the electron beam energies used for CL depth profiling. Inset: Depth of damage in single crystal (100) diamond as a function of oxygen ion energy determined using the data shown in (a) and (b). c) Depth distributions of oxygen implanted in amorphous carbon simulated using SRIM for ion energies of 2, 8, 16 and 30 keV.

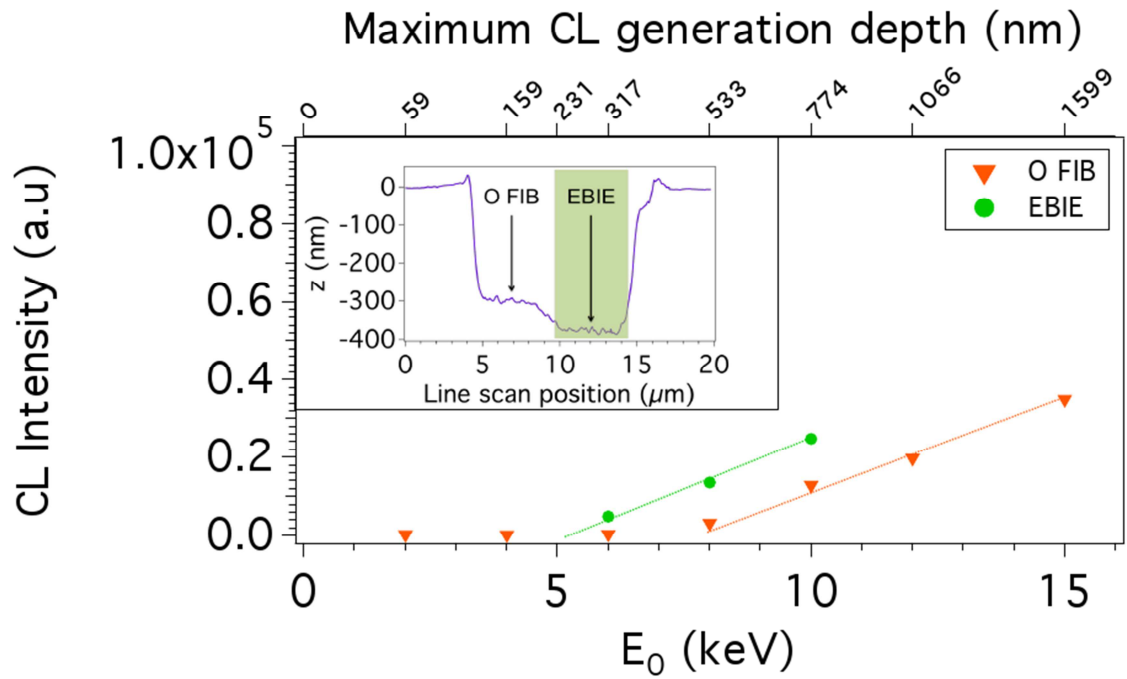

Figure 3. **Removal of damaged material by EBIE.** CL depth profile measured from a region milled by a 30 keV oxygen FIB before (`O FIB') and after EBIE was used to remove ~ 70 nm of surface material (`EBIE'). Inset: AFM line scan across measured region.